# Wave-Particle Fluctuations, Coherence,

# and Bose-Einstein Condensation

N. M. Chase
School of Arts and Sciences
Massachusetts College of Pharmacy and Health Sciences
Boston, MA   02115,   USA

email:   Norma.Chase@mcphs.edu

*Abstract*

By extending Einstein's separation of wave and particle parts of the second order thermal fluctuation to encompass "generalized fluctuations" in any Bose field, P. E. Gordon has proposed alternative definitions for nth order coherence and nth order coherent states; a state which is coherent to order n is one which contains no "wave" contributions to its number fluctuations $\langle (\Delta N)^m \rangle$, for all m ≤ n. This paper begins with a proof of the equivalence of Gordon's coherence conditions to those of Glauber; we then use the equivalent conditions to derive a one parameter family of states which are coherent to finite order. The main point of this paper is to explore some of the physical insights to be gained by extending dualism to higher orders. Recent experiments have examined aspects of the coherence of Bose-Einstein condensates. It has been argued that the condensate state is coherent to (at least) second or third order, but the coherence properties of Bose-Einstein condensates remain somewhat controversial. Using probability distributions developed by M. O. Scully and V. V. Kocharovsky et. al. , we apply Gordon's dualistic expression of the coherence conditions to investigate coherence properties in Bose-Einstein condensation (BEC). Via numerical calculations, we present a graphical survey of wave-like and particle-like fluctuations in condensed and uncondensed fractions. Near the critical point, we find a very marked peak in the ratio of nth order wave to nth order particle fluctuations. Not surprisingly, the n-point correlations between the positions of condensate atoms also peak near the critical temperature and this apparently mirrors, to all higher orders, the well-known relation between the integral of the 2-point correlation function over a certain volume and the rms fluctuation in the number of particles in that volume.

PACS 03.65 - Quantum theory, quantum mechanics

# 1 Introduction

Einstein's works [1] on the nature of the fluctuations $\langle (\Delta N)^2 \rangle = \langle N \rangle + \langle N \rangle^2$ in black-body radiation (1909) and in the molecular quantum gas (1924,1925) were turning points in the early development of quantum theory. For black-body radiation, he showed that while the second term would be expected for a system of electromagnetic waves, the first term would arise from a system of independent particles. This separation provided one of the first clear indications of the dual nature of light. In later works, Einstein extrapolated this same dualistic interpretation to the second order fluctuations in a molecular quantum gas. He noted that while first term would be expected for the familiar Poissonian fluctuation of (distinguishable) particles, the presence of the second term indicated that matter should demonstrate "novel" wave-like properties. In more recent times, Glauber [2] has given us the definitions of the fully coherent state and finite orders of coherence. A fully coherent state is the closest quantum analog of a "definite" classical wave; correlations reduce to Poissonian form to all orders. In a state which is coherent to only a finite order, correlations reduce to Poissonian form to all orders up to and including n, the order of coherence. Gordon [3] has proposed the extension of Einstein's separation of the second order wave and particle fluctuations in thermal radiation to encompass higher order fluctuations (moments about the mean) in any Bose field. In terms of this dualistic separation, Gordon proposed an alternative definition of the conditions for nth order coherence and demonstrated its validity up to 4$^{th}$ order. We begin this paper by proving the equivalence of Gordon's coherence conditions to those of Glauber. We then apply Gordon's dualistic representation of the coherence conditions to derive a one parameter family of states which are coherent to finite order [4,5]. With this background in place, we proceed to the main point of this paper, a first examination of some of the physical implications of extending dualism to higher orders.

Recent historic experiments have examined aspects of the coherence of Bose-Einstein condensates [6] . While it has been argued that the state of the condensate is coherent to (at least) second or third order, the higher order coherence properties of the condensate remain controversial [7-12]. M. O. Scully, Vl. V. Kocharovsky, and V. V. Kocharovsky [9-12] have developed non-equilibrium approaches to the dynamics and statistics of an ideal N-atom Bose gas cooling via interaction with a thermal reservoir and have thereby obtained equilibrium distributions for the number of atoms in the ground states of different model traps. Through a comprehensive examination of the equilibrium fluctuations and cumulants of the Bose-Einstein condensate, [12] shows that for an ideal Bose gas in any trap the distribution of noncondensed atoms becomes Poissonian (coherent) at very low temperatures while its "mirror image" distribution, that of the condensate, does not.

In this paper we use the probability distributions developed in [9,10] to explore the temperature dependence of coherence properties in BEC. Based on numerical calculations we present a graphical survey of wave-particle fluctuations in both condensate and noncondensate. Discussion is restricted to the quasithermal and low temperature approximations for the isotropic harmonic trap and homogeneous gas in a box models presented in [10]. In the quasithermal approximation of [10], we find a very marked peak in the ratio of nth order wave to nth order particle fluctuations near the critical temperature.. Not surprisingly, n-point correlations between the positions of condensate atoms likewise peak near the critical point; this result appears to mirror, to all higher orders, the well-known relation [18] between the integral of the 2-point correlation function over a certain volume and the rms fluctuation in the number of particles in that volume. We conclude by noting the relation between the results of this paper and the physical insights on the "loop gas" model of Matsubara [15] and Feynman [16] brought forth in papers by W. J. Mullin [17].
.

## 2. Background

2.1 Probability Distribution for nth Order Coherent States Derived from Glauber's Coherence Conditions

It is well known that a Bose state which is coherent to finite order n [2] satisfies the set of conditions

$$\langle :N^i: \rangle = \langle N \rangle^i \qquad \text{for all integers i in the interval } 2 \leq i \leq n, \qquad (1)$$

where N is the boson number operator $N = a^+ a$ and the dots denote normal ordering.

Stated otherwise,

$$\left\langle \left(a^+\right)^i a^i \right\rangle = \left\langle a^+ a \right\rangle^i \qquad \text{for all } 2 \leq i \leq n. \qquad (2)$$

A fully coherent state satisfies these conditions for all i up to infinity.

For a single mode expanded as a superposition of number eigenstates

$$|\phi\rangle = \sum_{N=0}^{N_{max}} C_N |N\rangle , \qquad (3)$$

it is straightforward to derive a one parameter family of states which are coherent to arbitrary finite order n [4].

The conditions for nth order coherence provide the equations

$$\sum_{N=i}^{N_{max}} P(n,N) \frac{N!}{(N-i)!} = \langle N \rangle^i \qquad 2 \leq i \leq n, \qquad (4)$$

where $P(n,N) = \left|C_N^{(n)}\right|^2$.

Combining (4) with the normalization condition

$$\sum_{N=0}^{N_{max}} P(n,N) = 1 \qquad (5)$$

and

$$\sum_{N=1}^{N_{max}} N\, P(n,N) = \langle N \rangle, \qquad (6)$$

and taking $N_{max}$ = n, we obtain

$$\begin{pmatrix} \langle N \rangle^0 \\ \langle N \rangle^1 \\ \langle N \rangle^2 \\ \langle N \rangle^3 \\ \langle N \rangle^4 \\ \vdots \\ \langle N \rangle^n \end{pmatrix} = \begin{pmatrix} 1 & 1 & 1 & 1 & 1 & \cdots & 1 \\ 0 & 1 & 2 & 3 & 4 & \cdots & n \\ 0 & 0 & \frac{2!}{0!} & \frac{3!}{1!} & \frac{4!}{2!} & \cdots & \frac{n!}{(n-2)!} \\ 0 & 0 & 0 & \frac{3!}{0!} & \frac{4!}{1!} & \cdots & \frac{n!}{(n-3)!} \\ 0 & 0 & 0 & 0 & \frac{4!}{0!} & \cdots & \frac{n!}{(n-4)!} \\ \vdots & \vdots & \vdots & \vdots & \vdots & \vdots & \vdots \\ 0 & 0 & 0 & 0 & 0 & 0 & \frac{n!}{0!} \end{pmatrix} \begin{pmatrix} P(n,0) \\ P(n,1) \\ P(n,2) \\ P(n,3) \\ P(n,4) \\ \vdots \\ P(n,n) \end{pmatrix} . \tag{7}$$

Multiplying both sides of (7) by the inverse of the matrix on the right hand side, we obtain

$$\begin{pmatrix} P(n,0) \\ P(n,1) \\ P(n,2) \\ P(n,3) \\ P(n,4) \\ \vdots \\ P(n,n) \end{pmatrix} = \begin{pmatrix} \frac{(-1)^0}{0!0!} & \frac{(-1)^1}{0!1!} & \frac{(-1)^2}{0!2!} & \frac{(-1)^3}{0!3!} & \frac{(-1)^4}{0!4!} & \cdots & \frac{(-1)^n}{0!n!} \\ 0 & \frac{(-1)^0}{1!0!} & \frac{(-1)^1}{1!1!} & \frac{(-1)^2}{1!2!} & \frac{(-1)^3}{1!3!} & \cdots & \frac{(-1)^{n-1}}{1!(n-1)!} \\ 0 & 0 & \frac{(-1)^0}{2!0!} & \frac{(-1)^1}{2!1!} & \frac{(-1)^2}{2!2!} & \cdots & \frac{(-1)^{n-2}}{2!(n-2)!} \\ 0 & 0 & 0 & \frac{(-1)^0}{3!0!} & \frac{(-1)^1}{3!1!} & \cdots & \frac{(-1)^{n-3}}{3!(n-3)!} \\ 0 & 0 & 0 & 0 & \frac{(-1)^0}{4!0!} & \cdots & \frac{(-1)^{n-4}}{4!(n-4)!} \\ \vdots & \vdots & \vdots & \vdots & \vdots & \vdots & \vdots \\ 0 & 0 & 0 & 0 & 0 & 0 & \frac{(-1)^0}{n!0!} \end{pmatrix} \begin{pmatrix} \langle N \rangle^0 \\ \langle N \rangle^1 \\ \langle N \rangle^2 \\ \langle N \rangle^3 \\ \langle N \rangle^4 \\ \vdots \\ \langle N \rangle^n \end{pmatrix} . \tag{8}$$

In more compact form,

$$P(n, \langle N \rangle, N) = \frac{\langle N \rangle^N}{N!} \sum_{j=0}^{n-N} \frac{(-\langle N \rangle)^j}{j!} . \tag{9a}$$

It is important to note that amongst the probabilities given in (9a) we find

$$P(n,\langle N \rangle, n-1) = \frac{\langle N \rangle^{n-1}}{(n-1)!}(1-\langle N \rangle),$$

If we restrict discussion to only positive probabilities, then the states (9a) are constrained to $0 < \langle N \rangle \leq 1$. In BEC, such states might conceivably represent the non-condensate (but not the condensate) at low temperatures.

For completeness we note that there also exists a set of states which are coherent to at most second order [5]

$$|II_{\max}\rangle = C_0|0\rangle + C_N|N\rangle \qquad N \geq 2$$

where
$$|C_0| = \sqrt{\frac{1}{N}}; \quad |C_N| = \sqrt{\frac{N-1}{N}}. \tag{9b}$$

These states differ from all other states which are coherent to finite order in that the average number of particles $\langle N \rangle = N-1$ may be arbitrarily large. Since $|C_0|^2 = \frac{1}{\langle N \rangle + 1}$ and $|C_N|^2 = \frac{\langle N \rangle}{\langle N \rangle + 1}$, we see that in the limit of very large $\langle N \rangle$ these states closely resemble number eigenstates.

## 2.2  Wave-Particle Duality and Combinatorial Aspects of Coherence

By extending Einstein's [1] separation of wave and particle parts of the thermal fluctuation $\langle (\Delta N)^2 \rangle = \langle N \rangle + \langle N \rangle^2$ to encompass generalized fluctuations $\langle (\Delta N)^m \rangle$ of any Bose field, Gordon [3] has proposed alternative definitions for nth order coherence and nth order coherent states. Noting that the fully coherent state is the closest quantum analog of a "definite" classical wave, Gordon proposed the following.

1) The mth order number fluctuation $\langle (\Delta N)^m \rangle$ in an arbitrary state may be separated into a sum of a Poissonian ("particle") term $\langle (\Delta N)^m \rangle_P$, given by the form of $\langle (\Delta N)^m \rangle$ for a fully coherent state, plus a "wave fluctuation" term $W_m$ given by

$$W_m = \langle (\Delta N)^m \rangle - \langle (\Delta N)^m \rangle_P \quad . \tag{10}$$

For the fully coherent state, wave fluctuations vanish to all orders.

2) A state which is coherent to order n is one which contains no "wave" contributions to its number fluctuations $\langle (\Delta N)^m \rangle$, for all m ≤ n, the order of coherence. That is, the generalized fluctuations in an nth order coherent state are strictly Poissonian ("particle-like") for all orders up to and including the order of coherence n.

$$\langle (\Delta N)^m \rangle = \langle (\Delta N)^m \rangle_P \quad \forall \ m \leq n. \tag{11}$$

It is straightforward to prove the equivalence of Gordon's (11) to Glauber's (1).

For an arbitrary state, the mth order fluctuation

$$\langle (\Delta N)^m \rangle = \sum_{N=0}^{N_{max}} (N - \langle N \rangle)^m \, P_N = \sum_{j=0}^{m} \left[ \frac{m!}{j!(m-j)!} \right] (-\langle N \rangle)^{m-j} \langle N^j \rangle, \tag{12}$$

can be expanded by using the normal ordered expansion of the boson number operator

$$\left( a^+ a \right)^j = \sum_{k=1}^{j} S(j,k) \left( a^+ \right)^k (a)^k \, ; \tag{13a}$$

the coefficients S(j,k) are the Stirling numbers of the second kind [13]

$$S(j,k) = \left( \sum_{N=1}^{k} \frac{N^j (-1)^{k-N}}{N!(k-N)!} \right) . \tag{13b}$$

The coherence conditions (2) thus imply that in a state coherent to finite order n, the expectation value $\langle N^j \rangle$ reduces to

$$\langle N^j \rangle = \sum_{k=1}^{j} S(j,k) \langle N \rangle^k \qquad \text{for all } j \leq n . \tag{14}$$

In the fully coherent state (14) of course holds for all $j \in [\, 1, \infty\, )$.

Now the sum over j in equation (12) runs up to m, so for any state which is coherent to an order $n \geq m$, (12) reduces to

$$\langle (\Delta N)^m \rangle = \sum_{j=0}^{m} \left[ \frac{m!}{j!(m-j)!} (-\langle N \rangle)^{m-j} \left( \sum_{k=1}^{j} S(j,k) \langle N \rangle^k \right) \right] . \tag{15a}$$

Equation (15a) clearly gives the mth order fluctuations for the fully coherent state, in which the order of coherence "n" is greater than any finite order of fluctuation m. Examples of (15a) in simplified form are given below.

$$\langle (\Delta N)^2 \rangle_P = \langle N \rangle \; ; \qquad \langle (\Delta N)^3 \rangle_P = \langle N \rangle ; \qquad \langle (\Delta N)^4 \rangle_P = \langle N \rangle (3\langle N \rangle + 1) ;$$

$$\langle (\Delta N)^5 \rangle_P = \langle N \rangle (10\langle N \rangle + 1); \qquad \langle (\Delta N)^6 \rangle_P = \langle N \rangle (15\langle N \rangle^2 + 25\langle N \rangle + 1); \tag{15b}$$

To see that (11) is equivalent to (2), note that since the Stirling numbers of the second kind S(j,k) are all positive, given (14) we could work our way back via (13a) to (2). (11) provides the necessary and sufficient conditions for (2).

This being the case, the probability distributions (9a) should be derivable from the pivotal restriction on moments about the origin given in (14). Combining (14) with

$$\langle N^j \rangle = \sum_{N=1}^{N_{max}} N^j \, P(N) \qquad (16)$$

and normalization $\sum_{N=0}^{N_{max}} P(N) = 1$, and taking Nmax = m, we obtain the m+1 equations

$$\begin{pmatrix} 1 & 0 & 0 & 0 & 0 & \cdots & 0 \\ 0 & S(1,1) & 0 & 0 & 0 & \cdots & 0 \\ 0 & S(2,1) & S(2,2) & 0 & \cdots & 0 \\ 0 & S(3,1) & S(3,2) & S(3,3) & 0 & \cdots & 0 \\ \vdots & S(4,1) & S(4,2) & S(4,3) & S(4,4) & & 0 \\ \vdots & \vdots & \vdots & \vdots & & & \vdots \\ 0 & S(m,1) & S(m,2) & S(m,3) & S(m,4) & \cdots & S(m,m) \end{pmatrix} \begin{pmatrix} \langle N \rangle^0 \\ \langle N \rangle^1 \\ \langle N \rangle^2 \\ \langle N \rangle^3 \\ \langle N \rangle^4 \\ \vdots \\ \langle N \rangle^m \end{pmatrix} = \begin{pmatrix} 1 & 1 & 1 & 1 & 1 & \cdots & 1 \\ 0 & 1^1 & 2^1 & 3^1 & 4^1 & \cdots & m^1 \\ 0 & 1^2 & 2^2 & 3^2 & 4^2 & \cdots & m^2 \\ 0 & 1^3 & 2^3 & 3^3 & 4^3 & \cdots & m^3 \\ 0 & 1^4 & 2^4 & 3^4 & 4^4 & \cdots & m^4 \\ 0 & \vdots & \vdots & \vdots & \vdots & & \vdots \\ 0 & 1^m & 2^m & 3^m & 4^m & \cdots & m^m \end{pmatrix} \begin{pmatrix} P(0) \\ P(1) \\ P(2) \\ P(3) \\ P(4) \\ \vdots \\ P(m) \end{pmatrix} \qquad (17)$$

Multiplying both sides of (17) by a matrix of Stirling numbers of the *first kind* collapses the LHS of (17) to the LHS of (7) with n = m; the RHS of (17) becomes equal to the RHS of (7), so we recover (9a).

Finally, we note that from (14) moments about the origin for the fully coherent state have forms such as

$$\langle N^2 \rangle_P = \langle N \rangle^2 + \langle N \rangle, \quad \langle N^3 \rangle_P = \langle N \rangle^3 + 3\langle N \rangle^2 + \langle N \rangle, \quad \langle N^4 \rangle_P = \langle N \rangle^4 + 6\langle N \rangle^3 + 7\langle N \rangle^2 + \langle N \rangle,$$

$$\langle N^5 \rangle_P = \langle N \rangle^5 + 10\langle N \rangle^4 + 25\langle N \rangle^3 + 15\langle N \rangle^2 + \langle N \rangle \quad , \quad \ldots \quad . \qquad (18)$$

(the jth moment about the origin contains powers of <N> up to "j"), while direct calculation of moments about the origin from (9a)

$$\left\langle N^j \right\rangle = \sum_{N=0}^{n} N^j P(n,N) = \sum_{N=0}^{n} N^j \left( \frac{\langle N \rangle^N}{N!} \sum_{i=0}^{n-N} \frac{(-\langle N \rangle)^i}{i!} \right) \tag{19}$$

would seem to invariably include powers of $\langle N \rangle$ greater than j. Thus, it is worthwhile to see how the collapse of (19) to (18) occurs for n ≥ j. After some algebra, equation (19) may be rewritten as

$$\left\langle N^j \right\rangle = \sum_{k=1}^{n} \left( \langle N \rangle^k \sum_{N=1}^{k} \frac{N^j (-1)^{k-N}}{N!(k-N)!} \right), \tag{20a}$$

and then the sum over k separated into two terms

$$\left\langle N^j \right\rangle = \sum_{k=1}^{j} \left( (\langle N \rangle)^k \sum_{N=1}^{k} \frac{N^j (-1)^{k-N}}{N!(k-N)!} \right) + \sum_{k=j+1}^{n} \left( (\langle N \rangle)^k \sum_{N=1}^{k} \frac{N^j (-1)^{k-N}}{N!(k-N)!} \right). \tag{20b}$$

Since

$$\sum_{N=1}^{k} \left( \frac{N^j (-1)^N}{N!(k-N)!} \right) = 0 \quad \text{for all k > j,} \quad [14] \tag{20c}$$

in (20b) every individual term in the second sum over k vanishes.

Thus, for all n ≥ j, (20b) reduces to

$$\left\langle N^j \right\rangle = \sum_{k=1}^{j} \left( (\langle N \rangle)^k \sum_{N=1}^{k} \frac{N^j (-1)^{k-N}}{N!(k-N)!} \right), \tag{21}$$

which is identical to (14) since the sums over N on the right hand side of (21) are the Stirling numbers of the second kind (13b).

We conclude this section by reviewing some physical implications. The significance of generalized fluctuations in N becomes apparent in higher order interference experiments which examine correlated coincidence rates in multiple detectors. It is well known that the left hand side of (2) is proportional to the correlated absorption rate in "i" detectors , while the absorption rate in a single detector is proportional to

$$\sum_f |\langle f|a|i\rangle|^2 = \langle i|a^+ a|i\rangle = \langle N\rangle \; ; \tag{22}$$

for a state coherent to order n, (2) implies that m-fold correlations are proportional to a product of absorption rates in the m detectors, for all m $\leq$ n the order of coherence. This suggests that the correlation is "accidental", as if one were detecting "independent" particles. Gordon's representation of (2) implies that the existence of any correlation *above* the "merely accidental" *requires the presence of wave-like contributions* to the generalized fluctuations in particle number. Specifically, for m-fold ('non-accidental") *correlation*, we must have *a wave-like* contribution to the mth order fluctuation,

$$\langle (\Delta N)^m \rangle \neq \langle (\Delta N)^m \rangle_P \; . \tag{23}$$

The association of the quantities $W_m$ in equation (10) with the quality of an "extended wave" thus seems physically reasonable.

   Let us return now to (13) and (14) and note their combinatorial content. The Stirling numbers of the second kind S(j,k) give the number of ways of partitioning a set of "j" objects into "k" non-empty sets. Stated otherwise, they give the number of ways of placing "j" distinguishable particles into "k" non-empty indistinguishable "bins".

For an arbitrary superposition of number eigenstates (not necessarily coherent to any order), (13a) implies

$$\langle N^j \rangle = \langle (a^+ a)^j \rangle = \sum_{k=1}^{j} S(j,k) \langle (a^+)^k (a)^k \rangle \; , \tag{24}$$

where $\langle N^j \rangle$ is proportional to the total coincidence rate in j detectors.

The "correlations" on the right hand side of (24) in general include "non-accidental" correlations (attributable to the presence of "wave-like" contributions to the fluctuation in particle number) as well as Poissonian, "particle-like" correlations.

$$\begin{pmatrix} \text{Total coincidence rate} \\ \text{in j detectors} \end{pmatrix} \propto \sum_{k=0}^{j} \begin{pmatrix} \text{Number of ways of placing j} \\ \text{distinguishable objects (particles)} \\ \text{into k non-empty "bins"} \end{pmatrix} \begin{pmatrix} \text{Correlated coincidence rate} \\ \text{in k detectors} \end{pmatrix} \quad (25)$$

However, for a state coherent to order $n \geq j$, with the "collapse" of (24) to the form (14), the total coincidence rate reduces to

$$\begin{pmatrix} \text{Total coincidence rate} \\ \text{in j detectors} \end{pmatrix} \propto \sum_{k=0}^{j} \begin{pmatrix} \text{Number of ways of placing j} \\ \text{distinguishable objects (particles)} \\ \text{into k non-empty "bins"} \end{pmatrix} \begin{pmatrix} \text{Product of the absorption rates} \\ \text{of k seemingly "independent"} \\ \text{particles} \end{pmatrix} . \quad (26)$$

Finally, we note that (11) implies that the sth cumulant of a state which is coherent to finite order reduces to Poissonian form for all $s \leq n$, the order of coherence; further, for $s \leq n + 2$, the sth cumulant differs from the Poissonian cumulant <N> by the state's sth order wave fluctuation.

## 3.  Application to Bose Einstein Condensation

We conclude by applying this dualistic approach to examine coherence properties in Bose-Einstein condensation. Based on probability distributions given in [9,10], we present a numerically generated graphical survey of temperature dependent wave-particle fluctuations in condensed and uncondensed fractions. Within the quasithermal and low temperature approximations, we examine fluctuations for a collection of atoms confined in an isotropic harmonic trap and for a homogeneous gas in a box.

The graphs were generated as follows. Using probability distributions in [10], we calculated moments about the mean and average numbers of particles; substituting the temperature-dependent <N>'s into (15b) we obtained Poissonian fluctuations. We found the corresponding "wave" fluctuations by then substituting into Equation (10).

## 3.1 Quasithermal Model for Atoms Confined in an Isotropic Harmonic Trap

### 3.1.1 In the Low Temperature Approximation

For a collection of 20 atoms confined in an isotropic harmonic trap, Figures 1 and 2 show second through fifth order fluctuations in condensate and noncondensate plotted versus the ratio T/Tc. In Figure 1 we see that at T = 0 condensate "wave" and "particle" fluctuations have equal absolute values. This result might have been anticipated given the limiting form of the condensate probability distribution given in [10]. As T approaches zero, the state of the condensate approaches the number eigenstate |Ncond> = |Natoms>. Since all moments about the mean vanish for a number eigenstate, each order of condensate wave fluctuation becomes as negative as the corresponding Poisson fluctuation becomes positive. From the Poissonian moments about the mean given in (15b), values at T = 0 are readily predicted for arbitrary number of atoms in the trap. At T = 0, second through fifth order Poissonian fluctuations have the values Natoms, Natoms, Natoms (3 Natoms+1), and Natoms (10 Natoms+1) respectively; corresponding wave fluctuations have values equal to minus one times each of these.

Figure 2 shows that as the system is cooled towards T = 0, wave fluctuations in the noncondensate decrease and approach zero and at extremely low temperature become negligible compared to Poissonian fluctuations. As the noncondensate becomes increasingly depleted, its state approaches finite orders of coherence. (This result is consistent with Figure 6, which shows noncondensate fluctuations in the "low temperature approximation" of [10].)

Comparing Figures 1 and 2 we note that, within the quasithermal model in the low temperature approximation [10], condensate fluctuations are sub-Poissonian while noncondensate are super-Poissonian.

### 3.1.2 In the High Temperature Approximation

We first note qualitative agreement of Figures 3 and 4 at very low temperatures with Figures 1 and 2. At higher temperatures new and interesting features appear.

Figure 3 shows that at higher temperatures, condensate wave fluctuations are positive to all orders ; fluctuations in the condensate become super-Poissonian. As the system is cooled towards Tc, wave fluctuations "peak" near the critical temperature. As the system is cooled still further, wave fluctuations in the condensate shrink to zero, though (notably) not all at the same temperature. At temperatures below Tc, condensate wave-like fluctuations are far more prominent than at very high T.

We note that plots of correlations between condensate atoms (not presented here) show similar behaviors. Correlations in the condensate are sub-Poissonian at low T, just as we have seen for condensate fluctuations. As the system is cooled towards Tc, wave-like contributions to condensate correlations peak at a temperature in the vicinity of Tc. At temperatures somewhat lower than this, wave-like contributions go markedly negative and become far more prominent than at very high temperatures.

Figure 4, not surprisingly, looks like a "reflection" of Figure 3. (Condensate and noncondensate probability distributions were, after all, constructed as "mirror images" of each other.) As the temperature is raised, we would anticipate a higher noncondensate fraction in the trap. For sufficiently high temperatures, the noncondensate approaches a number eigenstate |Nunc> = |Natoms>, with vanishing moments about the mean to all orders. At very high T, wave-particle fluctuations in the noncondensate have the same limiting values as the condensat at T = 0.

Finally, we note that plots of the wave-particle fluctuations for a homogeneous gas in a box (also not presented here) look quite similar to those we have shown for the isotropic harmonic trap, in both the quasithermal and "low temperature" approximations given in [10].

## 3.2   Low Temperature Approximation for Atoms in an Isotropic Harmonic Trap

Figures 5 and 6 display condensate and noncondensate fluctuations in the "low temperature approximation" given in [9,10]. Figure 5 shows condensate fluctuations which closely resemble those shown in Figure 1. In Figure 6 we see that, for a system containing as few as 20 atoms, noncondensate wave fluctuations very nearly vanish for an extended range of temperatures. (For orders 2 through 5, absolute values are all considerably smaller than $10^{-7}$ for T/Tc < 1 .) A similar result is obtained in the low temperature approximation for the homogeneous gas in a box, not displayed here. The noncondensate (to a good approximation) displays coherence to finite orders. As we show in an Appendix, this result might have been anticipated given the form of the underlying probability distribution given in [9,10].

## 4.   Conclusion

In the models presented in [10] and considered here, as the system is cooled, the state of the condensate apparently does not approach even finite orders of coherence; however, for temperatures near (and below) the critical temperature, "something" more prominently "wave-like" is going on within the condensate.

Due to space limitations, we have thus far presented only "toy systems" of 20 atoms. Of course one wonders about the behavior of larger numbers of atoms. A more direct comparison of wave and particle fluctuations might also be more satisfying. Figure 7 displays ratios of "wave-like" to "particle-like" fluctuations in the condensate

$$\frac{\left(\langle(\Delta N)^m\rangle - \langle(\Delta N)^m\rangle_P\right)}{\langle(\Delta N)^m\rangle_P}$$

(for m = 2, 3, 4, 5) in systems of 20, 50 and 100 atoms, in the high temperature approximation for the isotropic harmonic trap in the quasi-thermal model of [10]. (As T approaches zero, these ratios approach 1, and zero in the limit of very high T.) In Figure 7 we see that near the critical temperature in systems containing larger numbers of atoms, fluctuations in the condensate even more markedly display the qualities of an extended wave.

Not surprisingly, as shown in Figure 8, n-point correlations between the positions of condensate atoms also peak near the critical temperature. This result appears to mirror, to all higher orders, the well-known relation [18] between the integral of the 2-point correlation function over a certain volume and the second order fluctuation in the number of particles in that volume. In the notation of [18],

$$\int \nu dV = \frac{\langle (\Delta N)^2 \rangle}{\langle N \rangle} - 1. \tag{27}$$

The right hand side of (27) is, identically, the ratio of 2nd order wave fluctuations to 2nd order particle fluctuations.

Finally, let us consider the "loop gas" model of Matsubara [15] and Feynman [16], as illuminated in a beautiful paper by W. J. Mullin [17]. Mullin shows that as the system temperature approaches the critical point from above, there is an "explosion" in the mean number of particles associated with longer permutation loops. This is a direct consequence of the "overlap" of increasing numbers of condensate particles near the trap's center. The symmetry of the wave function under permutations of particles becomes increasingly crucial to its accurate description as one can no longer distinguish between different possible arrangements of an increasingly large number of identical particles. A phase transition occurs when the thermal de Broglie wavelength significantly exceeds the mean separation between atoms, so that there is a substantial overlap between thermal wavepackets, and this coincides with a catastrophic breakdown of the classical (distinguishable) "particle" description of the system. The explosion in the mean number of particles in large permutation loops occurs in the same narrow temperature range in which we find peaks in the ratio of generalized wave-like to particle-like fluctuations.

It will be interesting to investigate whether or not peaks in the ratios of generalized wave-like to particle-like fluctuations signal phase transitions in other systems. This question, and a more extensive study of the implications of a dualistic view of BEC (including for example atomic interactions [8], different model traps [10], and comparison with experimental data) remain for future work.

APPENDIX

In the low temperature approximation given in [10], the probability distribution for the number of condensed atoms $N_{cond}$ is given by

$$p_{N_{cond}} = \frac{1}{Z_{N_{atoms}}} \frac{H^{N_{atoms} - N_{cond}}}{(N_{atoms} - N_{cond})!}, \text{ with } Z_{N_{atoms}} = \exp(H) \frac{\Gamma(N_{atoms} + 1, H)}{N_{atoms}!} \quad (28)$$

and that for the number of uncondensed atoms $N_{unc}$

$$P_{N_{unc}} = p_{N_{atoms} - N_{unc}} = \frac{\exp(-H) N_{atoms}!}{\Gamma(N_{atoms} + 1, H)} \frac{H^{N_{unc}}}{N_{unc}!}, \quad (29)$$

Expressing the incomplete gamma function in (29) as

$$\Gamma(N_{atoms} + 1, H) = N_{atoms}! \exp(-H) \sum_{j=0}^{N_{atoms}} \frac{H^j}{j!}, \quad (30)$$

and rewriting (29) in the form

$$P_{N_{unc}} = \frac{H^{N_{unc}}}{N_{unc}!} \frac{1}{\sum_{j=0}^{N_{atoms}} \frac{H^j}{j!}}, \quad (31)$$

we note that at very low temperatures, for H < 1 and Natoms sufficiently large but finite, (31) is approximately equivalent to

$$P_{N_{unc}} = \frac{H^{N_{unc}}}{N_{unc}!} \sum_{j=0}^{N_{atoms}} \frac{(-H)^j}{j!}. \quad (32a)$$

In the small $N_{unc}$ regions of the distribution, where probabilities are "significant" for low T, (32a) is "almost indistinguishable" from

$$P_{N_{unc}} = \frac{H^{N_{unc}}}{N_{unc}!} \sum_{j=0}^{N_{atoms} - N_{unc}} \frac{(-H)^j}{j!} , \qquad (32b)$$

which (as shown in Section 2) is the probability distribution for a state which is coherent to an order $N_{atoms}$, with an average number of particles equal to H. We note that from (31) the average number of uncondensed atoms is given by

$$\langle N_{unc} \rangle = H \left( 1 - \frac{H^{N_{atoms}}}{N_{atoms}! \sum_{j=0}^{N_{atoms}} \frac{H^j}{j!}} \right) ; \qquad (33)$$

(33) implies that for H < 1 and large $N_{atoms}$, $\langle N_{unc} \rangle$ approaches H.

FIGURE CAPTIONS

Fig. 1. Condensate wave-particle fluctuations in the quasithermal model for a collection of 20 atoms confined in an isotropic harmonic trap in the low temperature approximation:
(a) $2^{nd}$ order; (b) $3^{rd}$ order; (c) $4^{th}$ order; (d) $5^{th}$ order.

Fig. 2. Non-condensate wave-particle fluctuations in the quasithermal model for a collection of 20 atoms confined in an isotropic harmonic trap in the low temperature approximation:
(a) $2^{nd}$ order; (b) $3^{rd}$ order; (c) $4^{th}$ order; (d) $5^{th}$ order. Insets show extreme low T behavior.

Fig. 3. Condensate wave-particle fluctuations in the quasithermal model for a collection of 20 atoms confined in an isotropic harmonic trap in the high temperature approximation:
(a) $2^{nd}$ order; (b) $3^{rd}$ order; (c) $4^{th}$ order; (d) $5^{th}$ order. Insets show extreme high T behavior.

Fig. 4. Noncondensate wave-particle fluctuations in the quasithermal model for a collection of 20 atoms confined in an isotropic harmonic trap in the high temperature approximation:
(a) $2^{nd}$ order; (b) $3^{rd}$ order; (c) $4^{th}$ order; (d) $5^{th}$ order.

Fig. 5. Condensate wave-particle fluctuations in the low temperature approximation for a collection of 20 atoms confined in an isotropic harmonic trap: (a) $2^{nd}$ order; (b) $3^{rd}$ order; (c) $4^{th}$ order; (d) $5^{th}$ order.

Fig. 6. Noncondensate wave-particle fluctuations in the low temperature approximation for a collection of 20 atoms confined in an isotropic harmonic trap: (a) $2^{nd}$ order; (b) $3^{rd}$ order; (c) $4^{th}$ order; (d) $5^{th}$ order.

Fig. 7.   Ratios of mth order wave-like to mth order Poissonian fluctuations in the condensate, for the quasi-thermal model for 20, 50 and 100 atoms in an isotropic harmonic trap, in the high T approximation:   (a) $2^{nd}$ order;  (b) $3^{rd}$ order;  (c) $4^{th}$ order;  (d) $5^{th}$ order.  Insets show extreme high T behavior.

Fig. 8.   n-point density correlations in the condensate, for the quasi-thermal model for 20, 50 and 100 atoms in an isotropic harmonic trap, in the high T approximation:   (a) n=2;  (b) n=3;  (c) n=4;  (d) n=5.

FIGURE 1    Condensate wave-particle fluctuations in the quasithermal model for a collection of 20 atoms confined in an isotropic harmonic trap in the low temperature approximation: (a) 2nd order; (b) 3rd order; (c) 4th order; (d) 5th order.

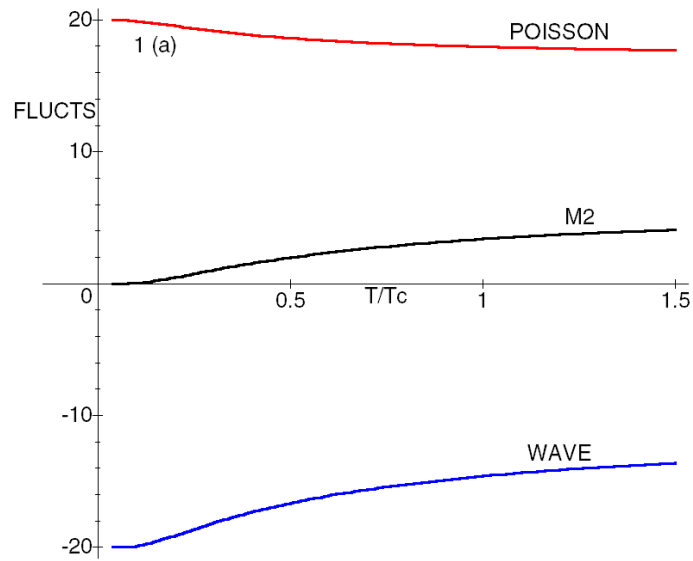
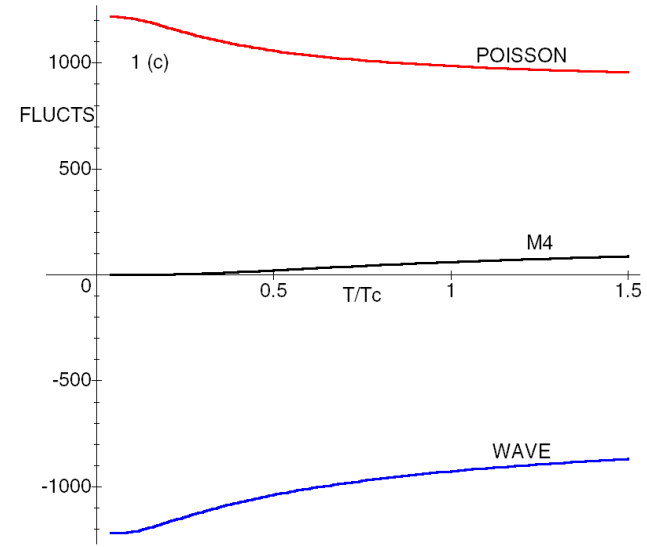
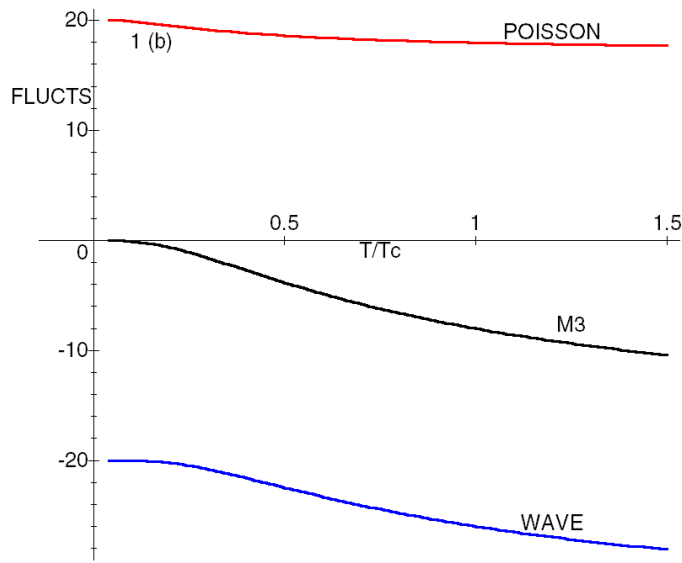
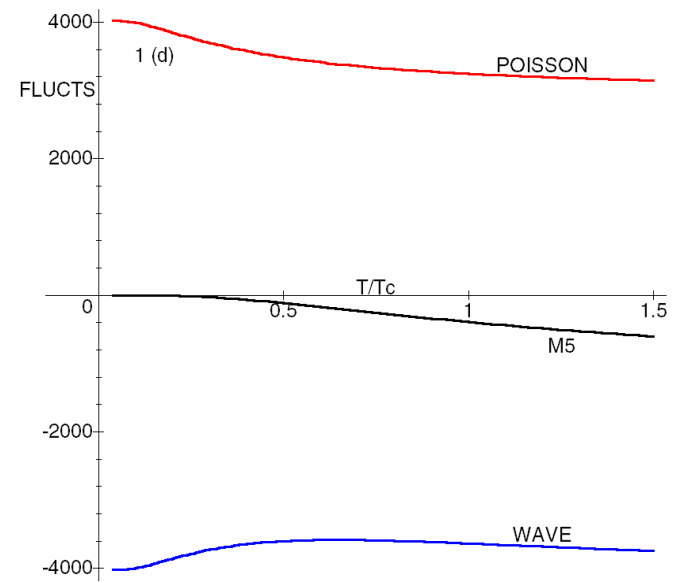

FIGURE 2 Non-condensate wave-particle fluctuations in the quasithermal model for a collection of 20 atoms confined in an isotropic harmonic trap in the low temperature approximation: (a) 2nd order; (b) 3rd order; (c) 4th order; (d) 5th order. Insets show extreme low T behavior.

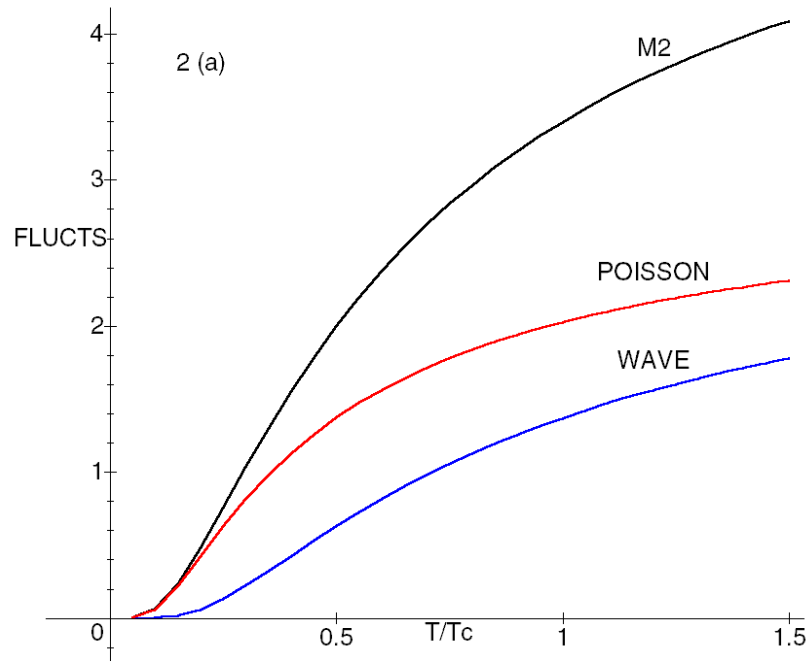
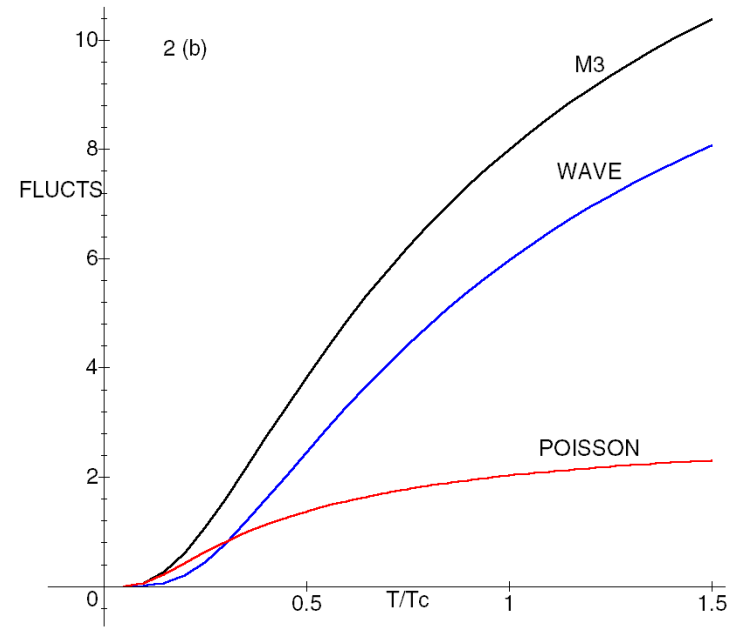
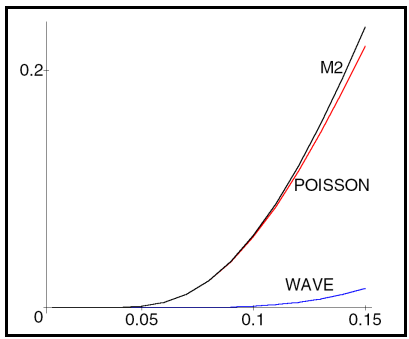
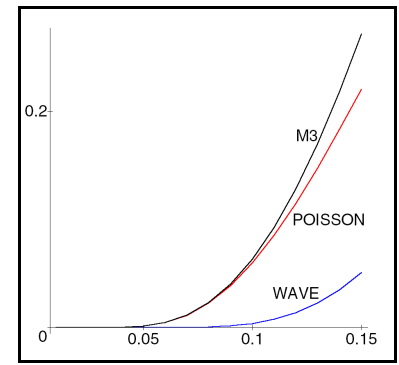

FIGURE 2 Non-condensate wave-particle fluctuations in the quasithermal model for a collection of 20 atoms confined in an isotropic harmonic trap in the low temperature approximation: (a) 2nd order; (b) 3rd order; (c) 4th order; (d) 5th order. Insets show extreme low T behavior.

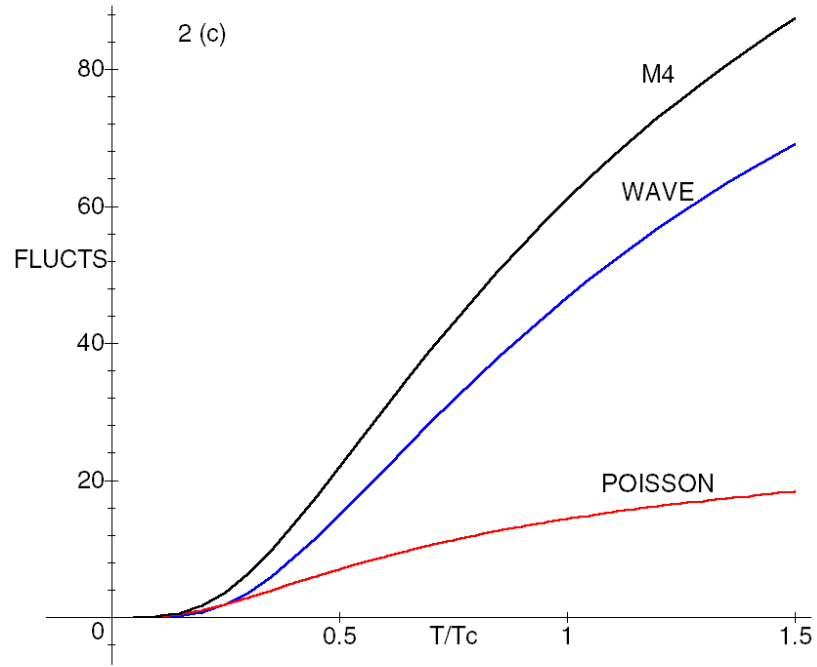
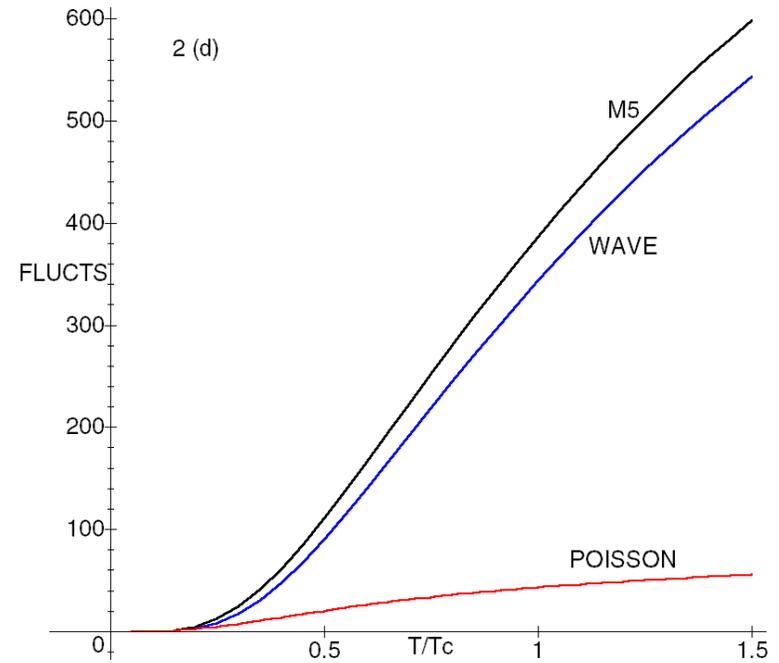
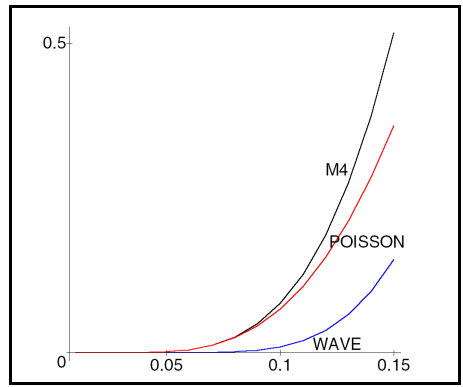
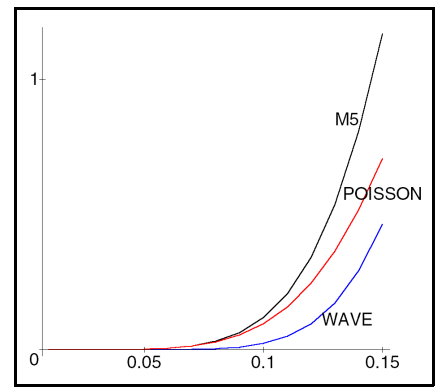

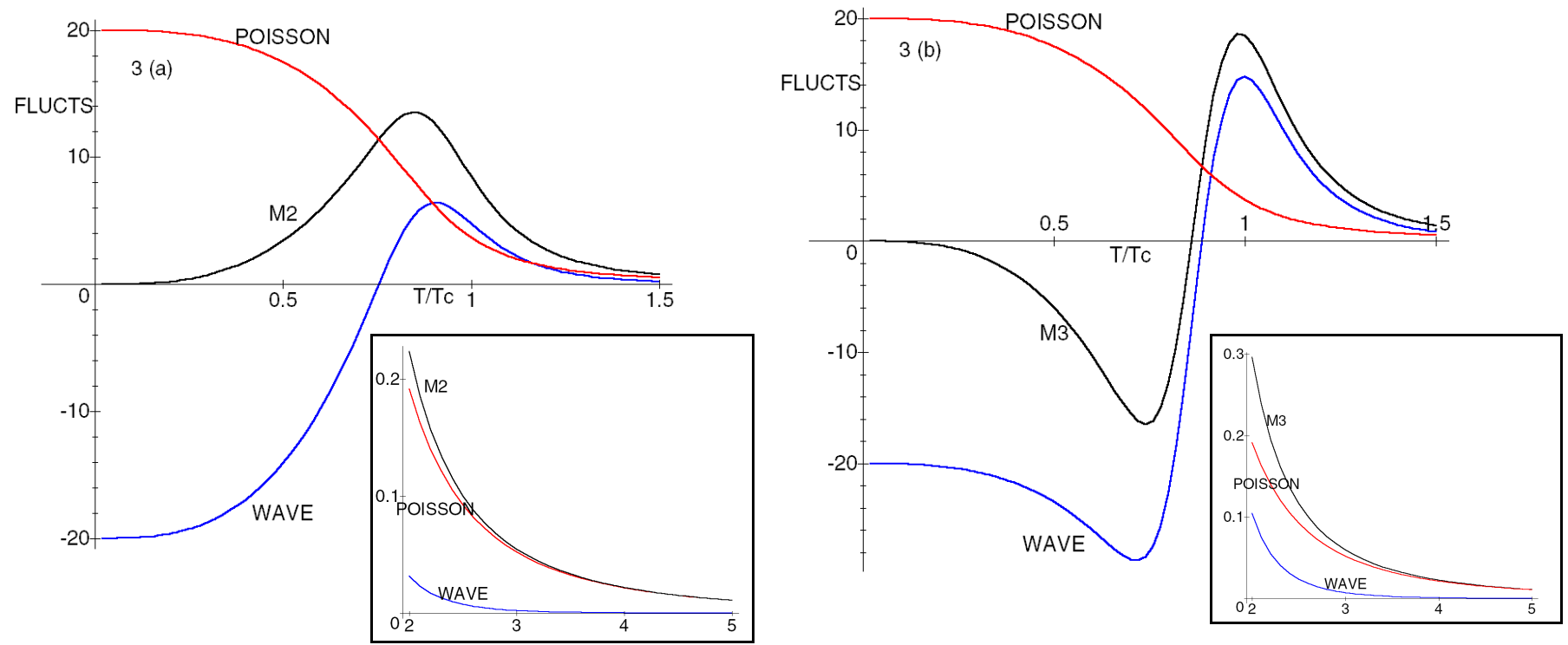

FIGURE 3  Condensate wave-particle fluctuations in the quasithermal model for a collection of 20 atoms confined in an isotropic harmonic trap in the high temperature approximation:  (a) 2nd order; (b) 3rd order; (c) 4th order; (d) 5th order.  Insets show extreme high T behavior.

FIGURE 3  Condensate wave-particle fluctuations in the quasithermal model for a collection of 20 atoms confined in an isotropic harmonic trap in the high temperature approximation: (a) 2nd order; (b) 3rd order; (c) 4th order; (d) 5th order. Insets show extreme high T behavior.

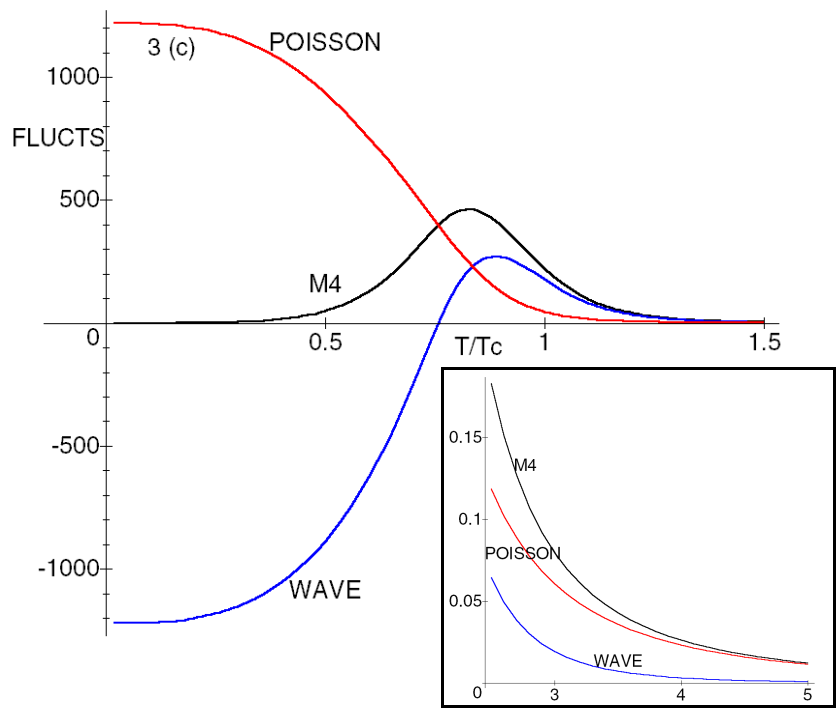
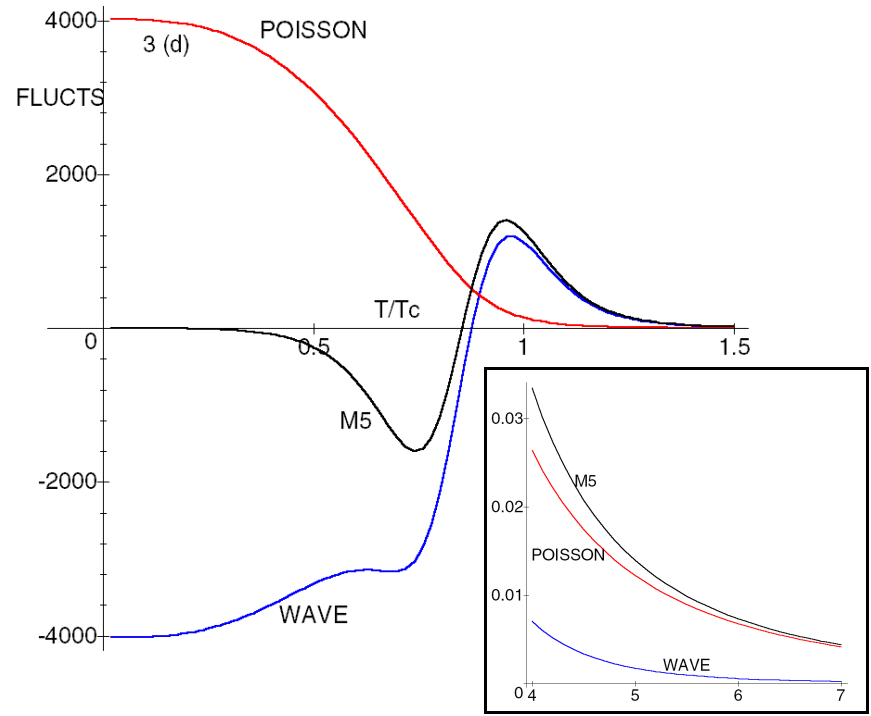

FIGURE 4    Noncondensate wave-particle fluctuations in the quasithermal model for a collection of 20 atoms confined in an isotropic harmonic trap in the high temperature approximation:  (a) 2$^{nd}$ order; (b) 3$^{rd}$ order; (c) 4$^{th}$ order; (d) 5$^{th}$ order.

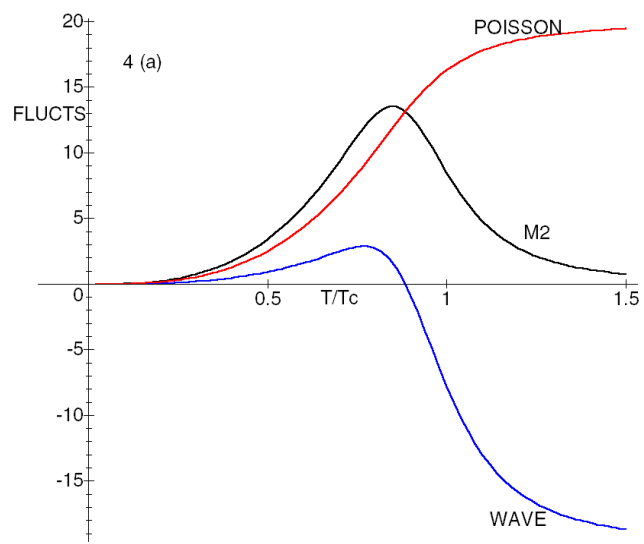
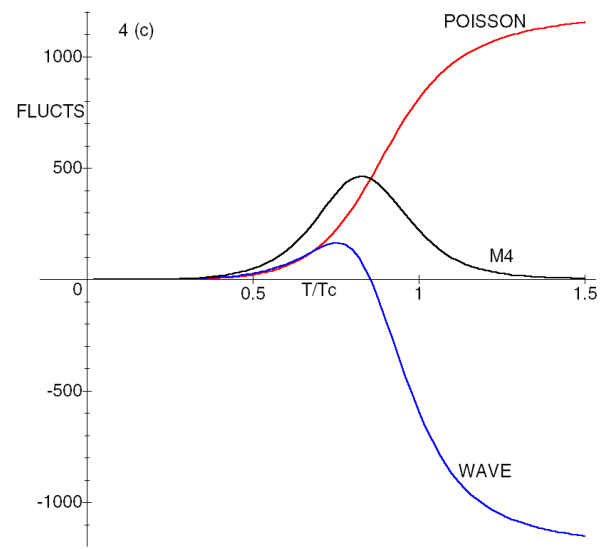
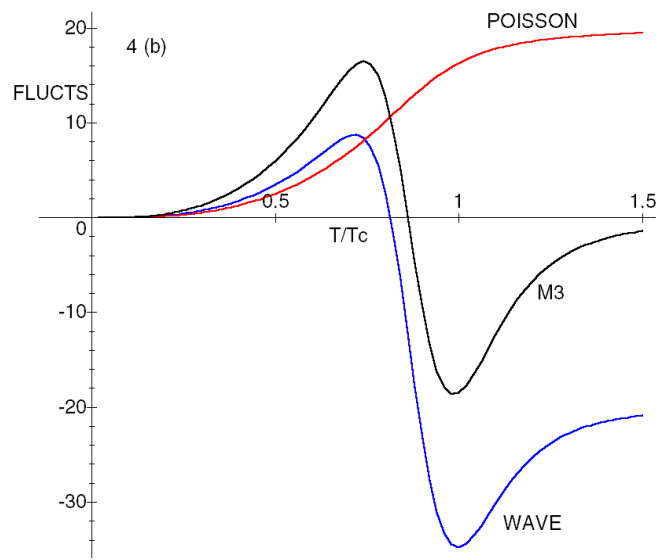
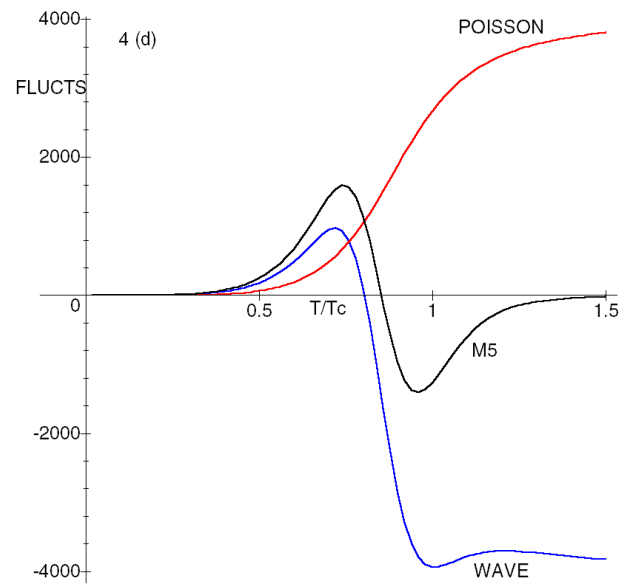

FIGURE 5  Condensate wave-particle fluctuations in the low temperature approximation for a collection of 20 atoms confined in an isotropic harmonic trap: (a) 2$^{nd}$ order; (b) 3$^{rd}$ order; (c) 4$^{th}$ order; (d) 5$^{th}$ order.

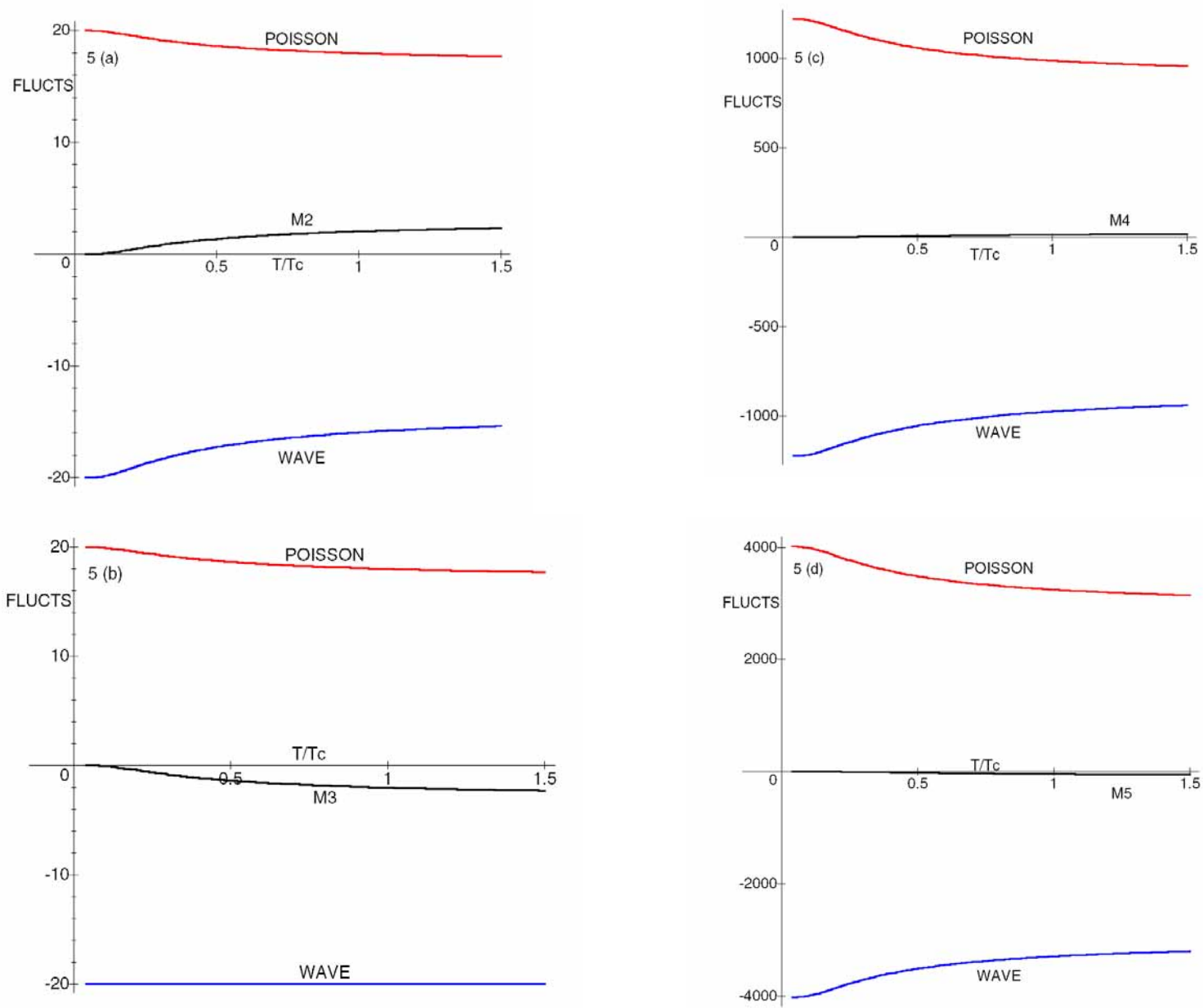

FIGURE 6  Noncondensate wave-particle fluctuations in the low temperature approximation for a collection of 20 atoms confined in an isotropic harmonic trap: (a) 2nd order; (b) 3rd order; (c) 4th order; (d) 5th order.

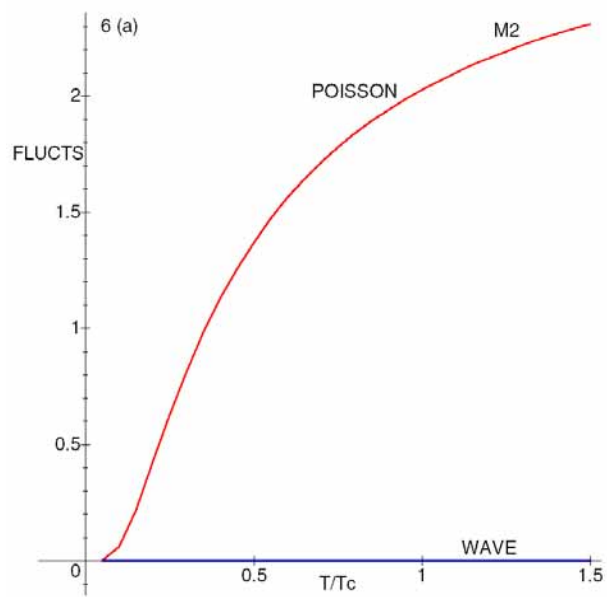
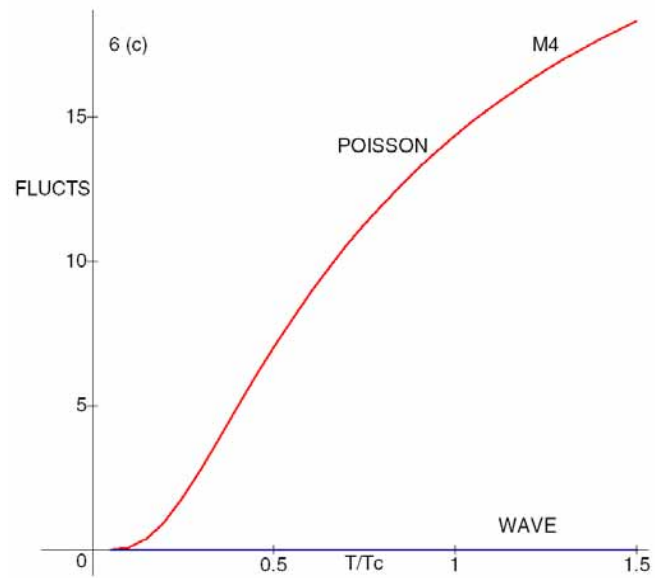
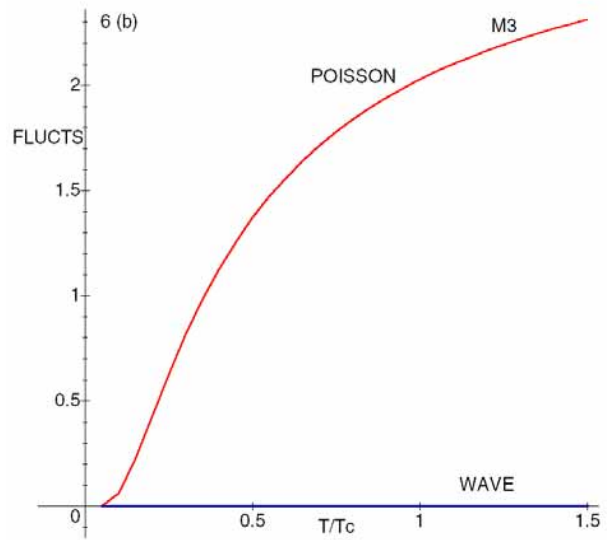
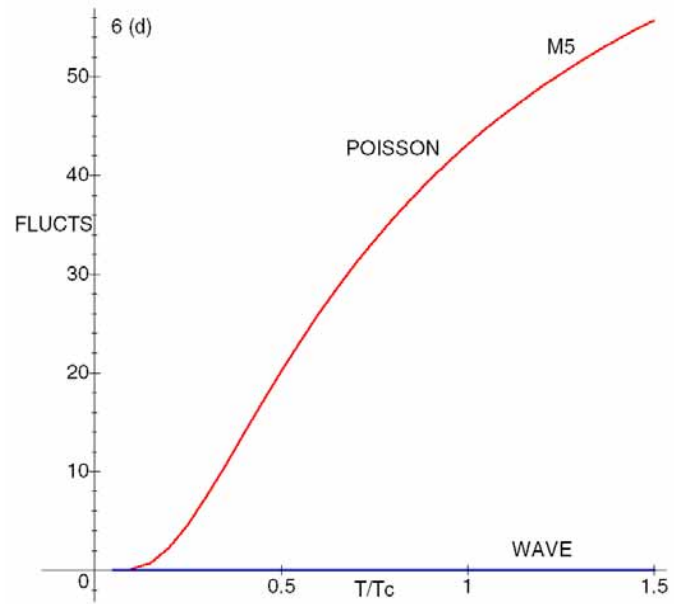

FIGURE 7: Ratios of mth order wave-like to mth order Poissonian fluctuations in the condensate, for the quasithermal model for 20, 50 and 100 atoms in an isotropic harmonic trap, in the high T approximation: a) 2nd order; (b) 3rd order; (c) 4th order; (d) 5th order. Insets show extreme high T behavior.

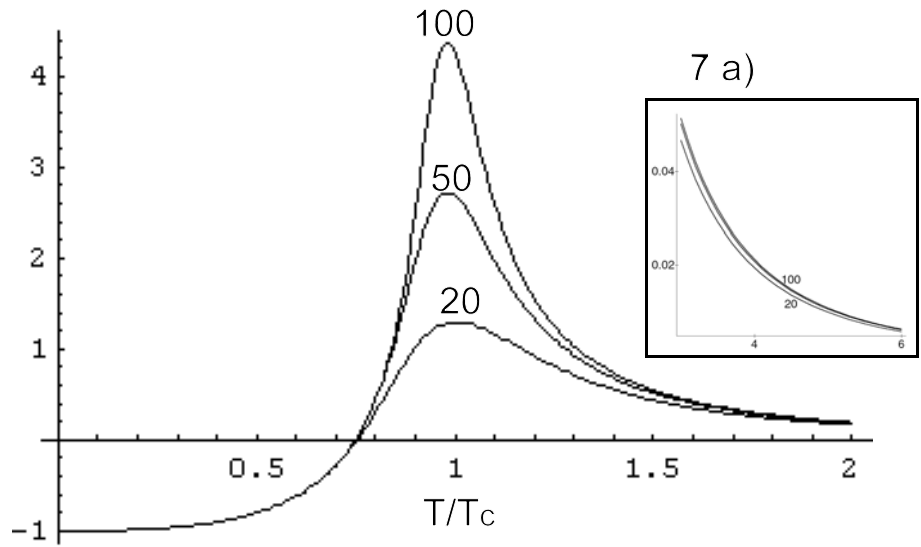
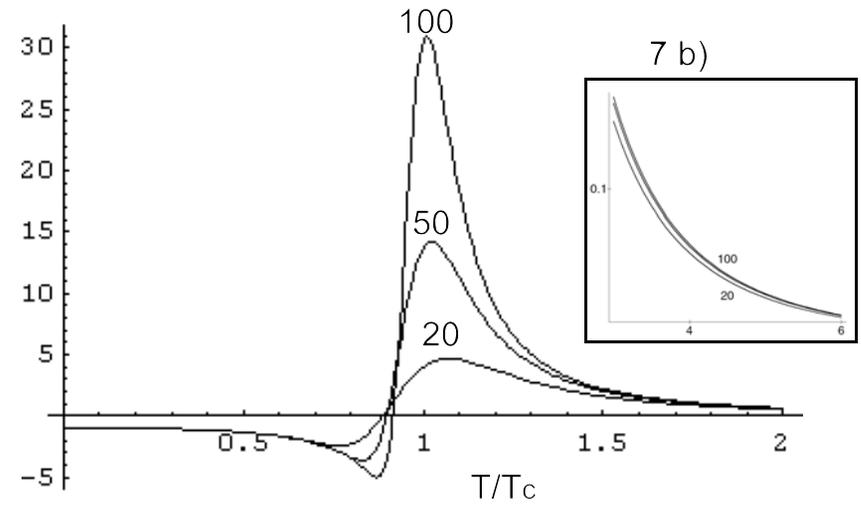

FIGURE 7: Ratios of mth order wave-like to mth order Poissonian fluctuations in the condensate, for the quasithermal model for 20, 50 and 100 atoms in an isotropic harmonic trap, in the high T approximation: a) 2nd order; (b) 3rd order; (c) 4th order; (d) 5th order. Insets show extreme high T behavior.

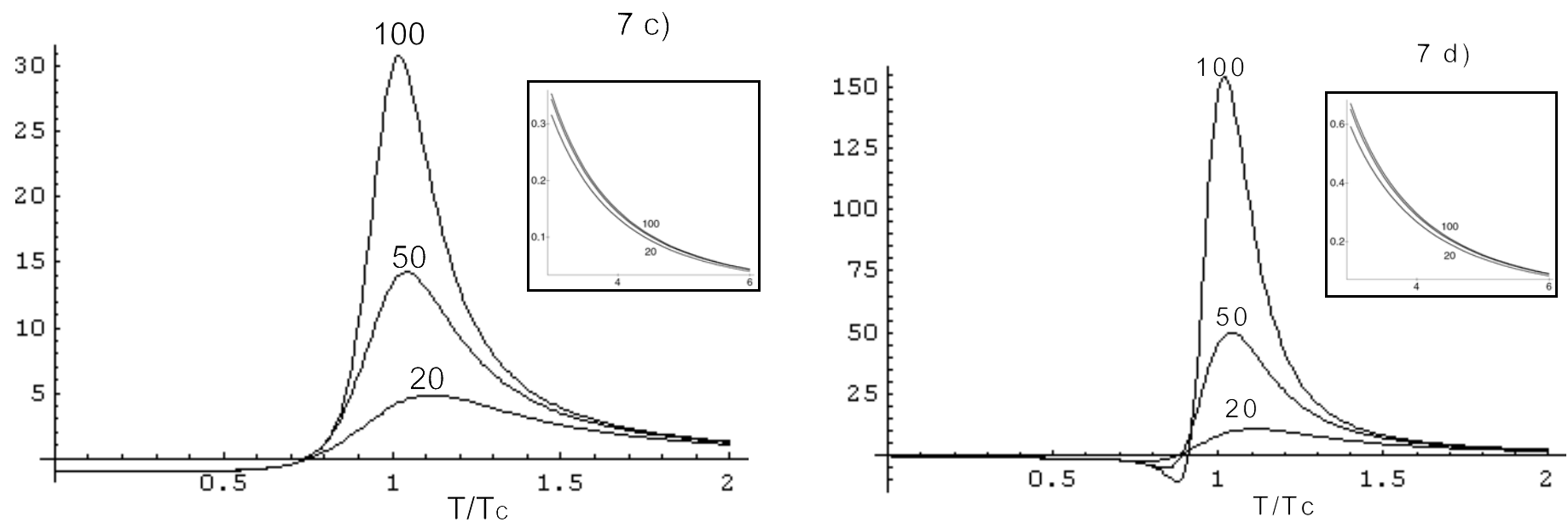

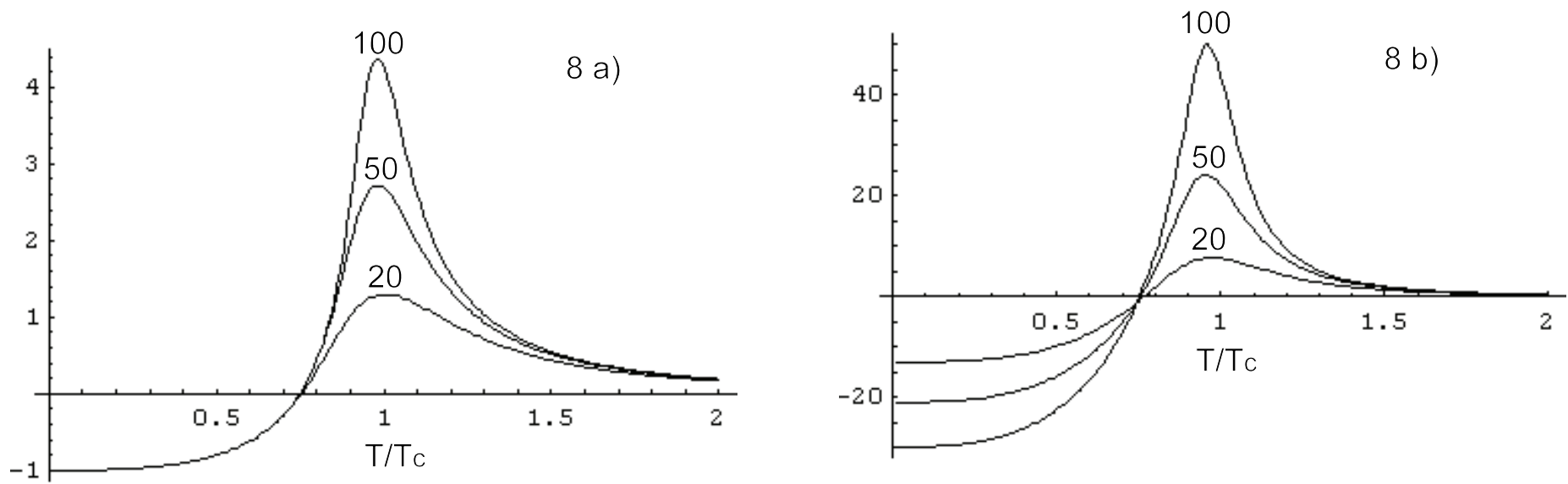

FIGURE 8: n-point density correlations in the condensate, for the quasi-thermal model for 20, 50 and 100 atoms in an isotropic harmonic trap, in the high T approximation: (a) n=2; (b) n=3; (c) n=4; (d) n=5.

FIGURE 8: n-point density correlations in the condensate, for the quasi-thermal model for 20, 50 and 100 atoms in an isotropic harmonic trap, in the high T approximation: (a) n=2; (b) n=3; (c) n=4; (d) n=5.

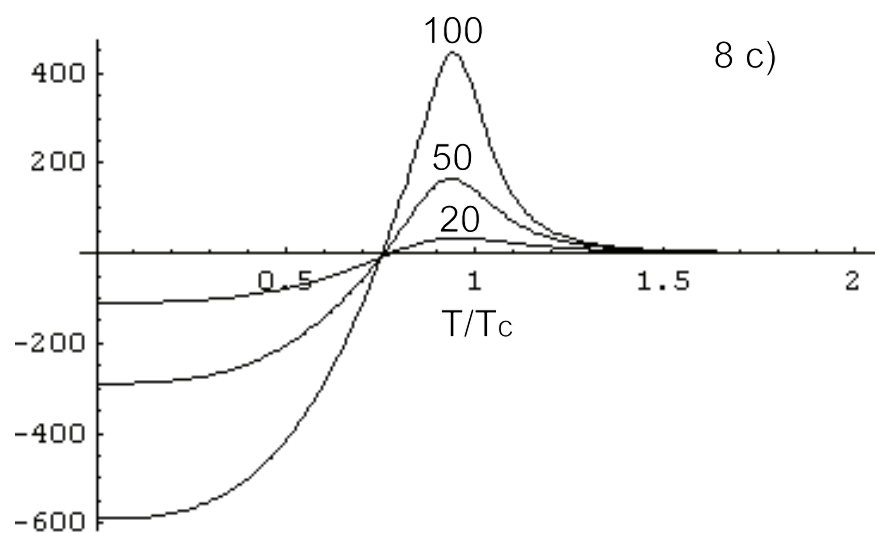
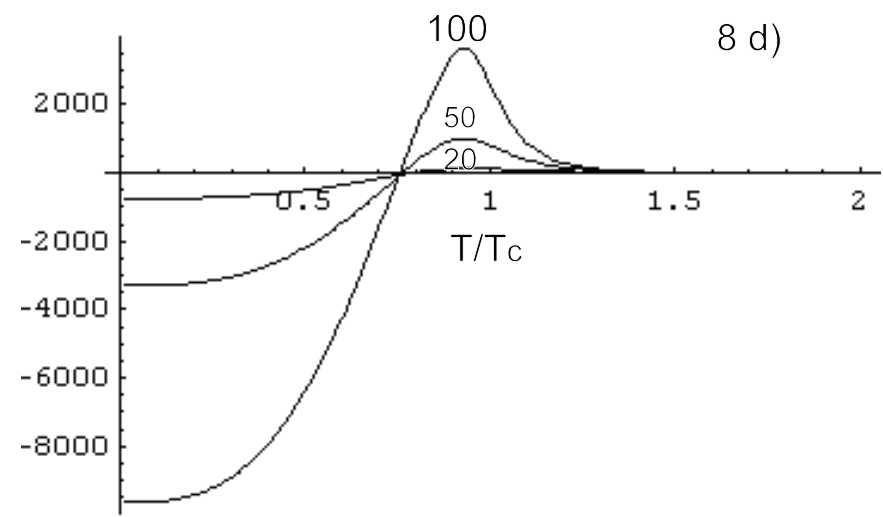